# ALPHAGMUT: A Rationale-Guided Alpha Shape Graph Neural Network to Evaluate Mutation Effects

**Boshen Wang**[1]     **Bowei Ye**[2]     **Lin Xu**[1*]     **Jie Liang**[2*]

[1] O'Donnell School of Public Health, University of Texas Southwestern Medical Center, Dallas, TX
Center for Bioinformatics and Quantitative Biology, Department of
[2] Biomedical Engineering, University of Illinois at Chicago, Chicago, IL
`{Boshen.Wang, Lin.Xu}@UTsouthwestern.edu`
`{boweiye2, jliang}@uic.edu`
* corresponding authors

## Abstract

*In silico* methods evaluating the mutation effects of missense mutations are providing an important approach for understanding mutations in personal genomes and identifying disease-relevant biomarkers. However, existing methods, including deep learning methods, heavily rely on sequence-aware information, and do not fully leverage the potential of available 3D structural information. In addition, these methods may exhibit an inability to predict mutations in domains difficult to formulate sequence-based embeddings. In this study, we introduce a novel rationale-guided graph neural network ALPHAGMUT to evaluate mutation effects and to distinguish pathogenic mutations from neutral mutations. We compute the alpha shapes of protein structures to obtain atomic-resolution edge connectivities and map them to an accurate residue-level graph representation. We then compute structural-, topological-, biophysical-, and sequence properties of the mutation sites, which are assigned as node attributes in the graph. These node attributes could effectively guide the graph neural network to learn the difference between pathogenic and neutral mutations using *k*-hop message passing with a short training period. We demonstrate that ALPHAGMUT outperforms state-of-the-art methods, including DEEPMIND's ALPHAMISSENSE, in many performance metrics. In addition, ALPHAGMUT has the advantage of performing well in alignment-free settings, which provides broader prediction coverage and better generalization compared to current methods requiring deep sequence-aware information.

## 1 Introduction

The missense mutation is one of the most frequent mutation types in the exon region of the human genome [1, 2, 3], which introduces the amino acid change in the translated protein. Many missense mutations have been found to alter protein functions and contribute towards diseases [4, 5, 6, 7, 8], while a large number of missense mutations are presumed to be functionally neutral [9, 10, 11]. Therefore, identifying potentially disease-causing missense mutations from passenger mutations is a critical task that can guide downstream analysis, such as comprehensively development of targeted therapy for cancer patients [12, 13, 14].

Previous studies mainly focus on utilizing sequence-aware information (e.g., alignment, co-evolution, language model, population genetics, a *priori* annotation, etc.) [15, 16, 17, 18, 19, 20, 21, 22, 23, 24], but do not leverage the rich biological features contained in protein structures. The rapid growth of experimental structures in the PROTEIN DATA BANK (PDB) and accurately predicted structures make it possible to incorporate structural information in predicting mutation effects [25, 26, 27]. A range of structure-based methods (ALPHAMISSENSE, DAMPRED, RHAPSODY, SPRI) have been developed to decipher structural properties, leading to improved mutation effect prediction [28, 29, 30, 31, 32].

However, existing methods face two notable limitations. First, protein function is usually collectively contributed by multiple residues in the domain or the structural scaffold [33, 34], as in the case of the catalytic region of the enzyme [35]. Hence, deciphering the impact from the neighboring environment of the mutation site is important along with the characteristics of the mutation itself. However, the majority of existing methods usually consider features of mutation individually and do not consider the impact of such neighboring residues, while a few of these methods utilize coarse-grained information (e.g., co-evolution, structural context, and contact profile) to encounter such influence [19, 28, 36, 30, 29, 32]. Therefore, a precise way to determine spatially relevant residues and formulate the impact of the neighboring environment is urgently needed. Second, many methods perform well on the benchmark datasets but report a large proportion (typically >40%) of mutations that are predicted as pathogenic in pan-cancer mutations or the human proteome [16, 29]. Such phenomena are contradictory to the common sense that the majority of mutations are functionally neutral [9, 10, 11, 37], which suggests these methods may exhibit high false positive rates in predicting unseen data. As a result, it is challenging to identify true pathogenic mutations in the noisy mutation landscape containing a high percentage of falsely predicted pathogenic mutations. Therefore, a method exhibiting higher specificity is required to reduce false positives.

In this study, we introduce a novel alpha-shape based graph model that generates an accurate residue-level graph representation of 3D protein structure, which is mapped from the atomic-resolution connection graph [38, 39]. We compute structural-, topological-, biophysical-, and sequence-aware features reflecting biological rationales of mutations and their spatially connected wild-type residues, which are assigned as node attributes. We then employed a graph neural network with $k$-hop message passing to account for neighborhood impact and mutation site simultaneously for predicting mutation effects [40, 41]. To further enhance the model's ability to learn accurate patterns of mutation effects, we implement multiple filtration steps to obtain likely cancer passenger mutations from those observed in real-world cancer patients, which provides a better ground truth of passenger mutations with functionally neutral effects.

Our novel method ALPHAGMUT achieves favorable overall performance compared with state-of-the-art methods including ALPHAMISSENSE from DEEPMIND, EVE, GMVP, and POLYPHEN-2. In addition, ALPHAGMUT exhibits the highest specificity which can significantly reduce false positives in analyzing real-world mutation data, as it is a common sense that the majority of mutations carry out neutral effects. Moreover, the structure-derived features capture essential biological rationales of mutations, which allows ALPHAGMUT to perform well in MSA-free setting. Hence, ALPHAGMUT holds better interpretability and generalization. It is also worth noting that ALPHAGMUT is capable of providing predictions for mutations located in protein domain with limited sequence-aware information, while EVE may suffer from the requirement of deep sequence-aware information and is unpredictable in these challenging domains.

## 2 Related Work

### 2.1 Mutation effect predictors

Previous methods for predicting mutation effects of missense mutations can be classified into four categories depending main features used in the model: alignment-based, co-evolution, structure-based, and language model approaches. Alignment-based methods are built based on protein-specific multiple sequence alignment (MSA), and then retrieve information about the mutation sites and local regions. MSA has proven effective in distinguishing pathogenic mutations from neutral ones, thus serving as a cornerstone and being combined with additional features (e.g., taxonomy distance, a *priori* annotation, allele frequency, secondary structures) in most methods [15, 16, 17, 18, 19, 20, 21, 22, 28, 36, 30, 32, 29]. Co-evolution methods further identify non-local residues along the linear MSA that likely have concurrent evolutionary relationships, suggesting spatial proximity and functional relevance to the mutation site [19, 24, 36]. Consequently, methods incorporating co-evolution signals show incremental performance [36, 19, 31]. Structure-based methods calculate structural, biophysical and topological properties based on experimental-determined structures or predicted structures, complementing sequence-aware information to make predictions [28, 30, 31, 32, 29]. Language models treat protein sequences as natural language and utilize transformer and other techniques to predict mutation effects according to sequence patterns [23].

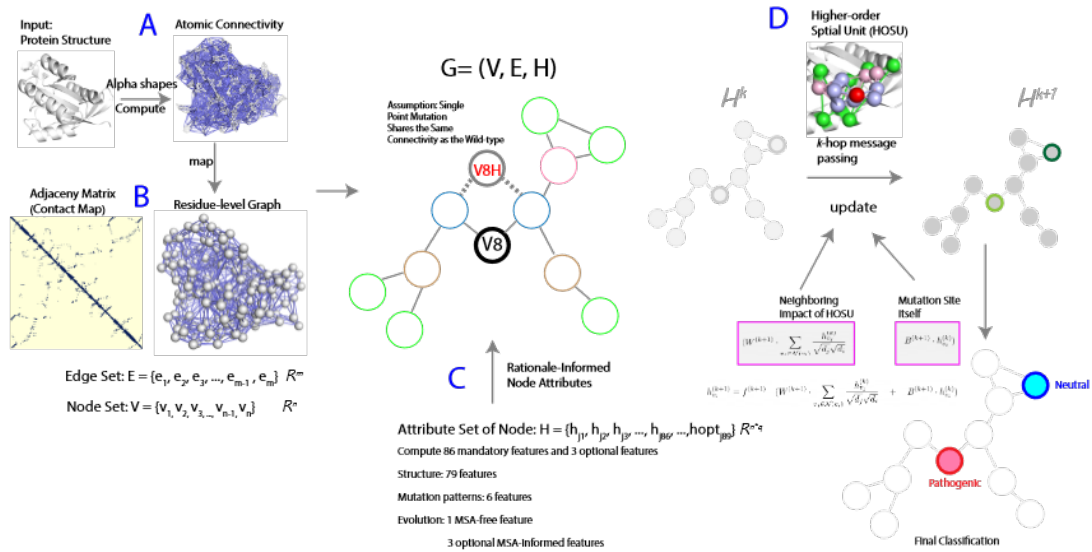

Figure 1: The architecture of the ALPHAGMUT. In the graph, a node represents the residue, and an edge represents the spatial connections between residues. We assume that the singular point mutation shares the same edge relationships as the wild-type graph. ALPHAGMUT provides node-level prediction, predicting whether a mutated node carries out a pathogenic or neutral effect. A) ALPHAGMUT computes the alpha shapes of 3D structure, obtains atomic connectivities, and maps to a residue-level graph. In the atomic-resolution graph, the blue edges connect atoms from different residues, and the grey edges connect atoms within the same residue. The blue atomic edges are used to identify residue-level connections. B) In the residue-level graph, each ball represents the alpha carbon of each residue, and blue edges connect residues with atomic connections indicated by the atomic-resolution graph. We then obtain the edge and node information in the graph, represented as $G = (V, E)$. The adjacency matrix (contact map) is also generated. C) ALPHAGMUT computes a wide range of features for the mutation site and its spatially connected residues. For each node, 79 structural-, biophysical, and topological features are computed based on 3D structure, six features describe static biophysics changes upon mutations, one sequence-aware feature (BLOSUM62 score) requiring no MSA. Three optional sequence-aware features are derived from MSA. For MSA-informed model, a total of 89 features reflect the biological rationales, which are assigned as node attributes. For MSA-free model, a total of 86 features are considered. We then obtain the full information of the graph as $G = (V, E, H)$. Raw features are then normalized. D) We utilize the graph convolutional network to learn patterns of mutation effects. A *k*-hop message-passing mechanism is employed to capture the neighboring impact for layered residues in close and distant regions of the mutation site. We use the $tanh$ as the activation function, and set $Adam$ function as the optimizer. The final output layer of the graph provides the probability indicating the likelihood of pathogenicity for a given mutation.

Previous studies have also employed a wide range of models, including random forest [22, 42, 30, 31, 32], Naive Bayes [16], Boltzmann machine [19], Gaussian mixture model [24], meta-analysis [43], graph attention neural network [36], transformer [23], and others. Among them, the majority of models are capable of learning patterns for each individual mutation site but lack the ability to take into account the impact of the spatially neighboring residues. A few methods are graph-based models to decipher such neighboring impact, but they exhibit limitations such as inadequate graph attributes reflecting biological rationales and coarse-grained graph representation leading to inaccurate connections among residues [36, 29, 28, 44].

Our method distinguishes itself from previous studies in two key aspects. First, ALPHAGMUT integrates a variety of well-informed features computed based on 3D structures, biophysical changes and sequence. These features complement each other, provide better model interpretability, and may enhance the performance. Such practice also provides the opportunity to make prediction without MSA information, which potentially achieves broader predictability for mutations located in domains without homologs (e.g., gap regions in MSA and orphan genes). Second, ALPHAGMUT provides an accurate means to identify neighboring residues with spatial proximity to the mutation site based on a novel graph representation. It further utilizes graph convolutional network (GCN) model with message passing mechanism to capture the neighboring impact of the mutation site.

## 2.2 Graph Representation of Protein

The graph representation of a protein is a key component in a graph-based model. Previous methods for generating the residue-level protein graph can be classified into three categories: co-evolution graph, centroid graph, and voxel graph. The co-evolution graph is a pseudo-graph that assumes residues sharing co-evolution relationships also share spatial proximity and functional relevance [36]. However, co-evolution may contain unrealistic connections as it is inferred from statistical analysis of MSA and other assumptions (e.g., pre-defined length of flanking regions of protein context) [36, 45, 46]. Additionally, it cannot illustrate detailed properties of 3D structures. The centroid graph only considers the centroid (*e.g.* alpha carbon) of each residue and disregards the complex spatial patterns harnessed by the folding arrangement and geometric shapes formed by side-chain and other backbone atoms. Due to its simplicity, it is the most widely used graph representation to determine residue-level connection [28, 44, 29, 47]. The voxel graph utilizes volume pixels to partition the 3D space of the protein structure and can provide both atomic-resolution and residue-level information. However, due to varying radii of different atom types, the fixed dimension size of cubic voxels is not able to perfectly represent spherical atoms [48].

Our method distinguishes itself from previous studies in two key aspects. First, ALPHAGMUT introduces a novel alpha-shape graph. It computes the meticulous atomic-resolution alpha shapes of the protein structure, and map it to a residue-level graph. Such graph representation captures accurate connectivity for any pairwise residues. Second, ALPHAGMUT formulates the higher-order spatial unit (HOSU), which provides a precise way to identify the relative distances of neighboring residues to the mutation site. HOSU plays an important role in the *k*-hop message passing process to account for the neighboring impact of the mutation site, which provide richer information than conventional one-layer message passing.

## 3 Methods

### 3.1 Generating Alpha-Shape Graph

ALPHAGMUT constructs a novel residue-level graph representation for a given high-quality 3D structure. It computes the alpha shapes which connect neighboring atoms at an alpha value as of $1.4\,\text{Å}$ (the radius of a water molecule) [38, 39, 49, 50], and then maps to a residue-level graph based on the connection information of atoms located among pairwise residues. Next, we obtain the edge and node information of the graph, which can be represented as

$$G = (V, E), \quad V \in \mathbb{R}^n, \quad E \in \mathbb{R}^m \tag{1}$$

$V$ is the set of nodes (residues), and $n$ is the number of nodes. $E$ is the set of edges (residue-level connections), and $m$ is the number of edges, with $m \leq \frac{n*(n-1)}{2}$. In this study, we consider the edges to have a bidirectional (undirected) relationship. The residue-level graph is a higher-order representation of atomic-resolution graph. Together they grant that our novel residue-level graph is SE3 equivariant.

Based on the residue-level graph, we then implement a Breadth-first search (BFS) algorithm to define the higher-order spatial unit (HOSU) of the mutation site, which captures residues with layered spatial proximity to the mutation site. We set the mutation site as the centre node, these residues with direct edges to the centre node are labeled as first-layer contacts. Second-layer contacts are those residues have direct edges to the first-layer contacts but without edges to the centre node. Following the same rule, we also obtain third-layer and forth-layer contacts. All residues identified in such searching process are grouped into the (HOSU) of the mutation site. The newly defined HOSU play roles in both graph attribute and message passing process.

## 3.2 Computing Biological Rationale as Graph Attributes

From the 3D structure, ALPHAGMUT computes features indicating atomic interactions, layered residue contact profiles, the geometric shape of local regions, and biophysical properties. The alpha shapes of the atomic graph provide detailed atomic interactions between the mutation site and its first-layer residues. We compute the count of each of the 16 types of atomic interactions with donor/acceptor effects, namely, $\{N_{CC}, N_{CN}, N_{CO}, N_{CS}, N_{NC}, N_{NN}, N_{NO}, N_{NS}, N_{OC}, N_{ON}, N_{OO}, N_{OS}, N_{SC}, N_{SN}, N_{SO}, N_{SS}\}$, which are based on atomic contact of heavy atoms (Carbon(C), Nitrogen(N), Oxygen(O), Sulfur(S)). We also compute the solvent accessible surface area, salt bridge, local geometric shape (buried, pocket, surface) of each residue. We retrieve three layered residue contact profile based on the newly built HOSU, which can be represented as $\{N_{Ala_{L1}}, N_{Ala_{L2}}, N_{Ala_{L3}}, N_{Arg_{L1}}, N_{Arg_{L2}}, N_{Arg_{L3}}, \cdots, N_{Val_{L1}}, N_{val_{L2}}, N_{Val_{L3}}\}$, and in total of 60 features are informed.

We also computes the static biophysical change upon the mutations, including changes of amino acid polarity, side-chain carbon-, oxygen-, nitrogen-, and sulfur changes, and backbone carbon change. BLOSUM62 substitution score servers as the MSA-free sequence-aware feature. We follow classic methods to build protein-specific MSA using homologs identified in UNIPROT eukaryotes reference proteomes, and obtain the site-specific entropy, wild-type frequency, and mutated-type frequency, and these three MSA-informed features are optional in the model.

We then vectorize all these features and concatenate them as the node attribute, and then obtain the full information of the graph

$$G = (V, E, H), V \in \mathbb{R}^n, E \in \mathbb{R}^m, H \in \mathbb{R}^{n*q} \tag{2}$$

The $H$ is the feature matrix of node attribute, same number size as the nodes, and in $q$ feature dimensions. For the MSA-informed model, $q = 89$. For the MSA-free model, $q = 86$.

## 3.3 Graph Neural Network

We employ a graph convolution network to learn the patterns of mutation effects. The message passing allows the model to update the latent representation of feature matrix $H$ incorporating the neighboring impact of spatially contact nodes and the mutation node simultaneously.

$$h_{v_i}^{(k+1)} = f^{(k+1)}(W^{(k+1)} \cdot \sum_{v_j \in \mathcal{N}(v_i)} \frac{h_{v_j}^{(k)}}{\sqrt{d_j}\sqrt{d_i}} + B^{(k+1)} \cdot h_{v_i}^{(k)}) \tag{3}$$

$\mathcal{N}(v_i)$ is the set of neighboring nodes connected to the target (point mutation) node $i$, $v_i$ is the target node, $v_j$ is the neighboring node, $d_i$ is the number of edges in the target node $i$, $d_j$ is the number of edges in the neighboring node $j$, $f^k, W^k, B^k$ are learnable parameters. $W^{(k+1)} \cdot \sum_{v_j \in \mathcal{N}(v_i)} \frac{h_{v_j}^{(k)}}{\sqrt{d_j}\sqrt{d_i}}$ is the explicit neighboring impact. $B^{(k+1)} \cdot h_{v_i}^{(k)}$ is the characteristic of the mutation site. Initial $h_{v_i}^{(0)}$ is the raw input feature vector of target node $i$.

We choose $k = 4$ convolution layers, which captures the information expanded to the fourth layers nodes. We set $tanh$ as the activation function, $Adam$ function as the optimizer, assign a random seed, learning rate as of $lr = 0.0001$, and weight decay as of $0.0005$. Early stopping is set as there is no improvement on the validation dataset and patience for 10,000 epochs, and we choose the epoch with the lowest validation loss to reduce overfitting. GCN is implemented with PYTORCH and PYTORCH GEOMETRIC, alpha-shape graph and features are implemented in PYTHON and C.

ALPHAGMUT captures essential distinguishing features as the initial inputs and shows great computation efficiency and feasibility, as a typical training period on one fold takes about 5 minutes in a workstation equipped with a single NVIDIA 3090TI and AMD RYZEN 7900X.

## 3.4 Benchmark Dataset

To guide the model to learn accurate patterns and obtain an unbiased comparison, we curate a novel benchmark dataset with rigorous criteria. We require the proteins to have high-quality PDB structures and disregard predicted structures, since 38.1% of residues in human proteomes have low-confidence predictions from ALPHAFOLD2, which may lead to inaccurate structure-derived graphs and features. We then consider mutations that have strong evidence to have pathogenic and neutral effects from both Mendelian diseases and cancer. We include a recently available dataset UNIFYPDBACCEPTABLE dataset reflecting mutation effects in Mendelian disorders [32], which selects high-confidence genes and mutations based on widely-accepted HUMVAR, HUMDIV and PREDICTSNP datasets [16, 42, 30, 31].

Table 1: Averaged 10-fold Cross-validation Performance Metrics

|  | ALPHAGMUT MSA-informed | ALPHAGMUT MSA-free | ALPHAMISSENSE | EVE (biased ratio) | GMVP | POLYPHEN-2 |
|---|---|---|---|---|---|---|
| AUPRC | **0.878** | 0.869 | 0.770 | 0.910 | 0.815 | 0.845 |
| AUROC | **0.886** | 0.877 | 0.823 | 0.892 | 0.831 | 0.847 |
| Recall | 0.836 | 0.832 | 0.789 | 0.831 | 0.871 | **0.884** |
| Specificity | **0.824** | 0.814 | 0.636 | **0.826** | 0.632 | 0.659 |
| Precision | **0.827** | 0.819 | 0.717 | 0.877 | 0.724 | 0.719 |
| Accuracy | **0.830** | 0.823 | 0.758 | 0.829 | 0.757 | 0.771 |
| F-1 | **0.831** | 0.825 | 0.788 | 0.850 | 0.790 | 0.793 |
| MCC | **0.661** | 0.647 | 0.576 | 0.649 | 0.521 | 0.556 |
| Completeness | 1 | 1 | 0.982 | 0.690 | 0.908 | 0.981 |

For mutations in cancer, we take CANCER MUTATION CENSUS (CMC) mutations by COSMIC as additional pathogenic mutations [51], which have clinical-, *in silico*, and experimental evidence. We then curate likely cancer passenger mutations by analyzing real-world mutation data in cancer patients. We focus on over 2.9 million pan-cancer missense mutations recorded by COSMIC v95 [51]. We select putative neutral genes with no record as cancer genes listed in ONCOKB [52], a lower gene-level mutation recurrence rate, no hyper-mutated mutation site, and no structural hotspot with multiple mutations [53]. Then mutations harbored by selected genes are additionally neutral mutations. While the majority of methods select neutral mutations from the general population using allele frequency thresholds [29, 21, 17], which may not necessarily exist in cancer genomes.

In total, our new benchmark dataset contains 5,715 pathogenic mutations and 4,852 neutral mutations among 469 proteins and 537 PDB chains. (Note: a singular protein may have several non-overlapped partial PDB structural domains (graphs) [54]). The 537 graphs contains a total of 183,727 nodes (173,160 wild-type nodes and 10,567 mutated nodes) and 1,008,656 edges. The mutated node are reasonably assumed to have same edge relationships, since backbone atoms are not changed (except beta-carbon missing in Gly) and backbones are rigid to form structural conformations [55, 56, 57, 58].

Due to only 29% of human proteins having high-quality PDB structures, our dataset contains the lowest number of mutations among methods in comparison. Therefore, we generate modified 10-fold cross-validation datasets. We random shuffle the index of each mutation node with ground truth, and select a balanced test dataset that includes 200 pathogenic and 200 neutral mutations for each-fold, all remaining nodes are in the training (90%) and validation (10%) process.

## 4 Results

### 4.1 Better Performance

We compare ALPHAGMUT with several state-of-the-art methods including structure-based deep learning method ALPHAMISSENSE by DEEPMIND, evolutionary generative model EVE, co-evolution graph neural network model GMVP, and the widely-used alignment method POLYPHEN-2. We re-align the residue index to reconcile a small proportion of different protein isoforms used by these compared methods. Binary decisions are made using their suggested thresholds, except EVE does not provide an universal threshold. We then choose the optimal pathogenicity likelihood $\theta$ as 0.37 for EVE to maximize its MCC in our dataset.

We comprehensively compare a variety of performance metrics, including the area under the ROC curve (AUROC), the area under the precision-recall curve (AUPRC), Matthews correlation coefficient (MCC), F-1 score, recall, specificity and accuracy. Detailed metrics are reported in the table 1 1. The ALPHAGMUT MSA-informed model achieves the overall best performance indicated by AUROC, AUPRC, MCC, F-1 score, specificity, accuracy, precision. The ALPHAGMUT MSA-free model also provides the same on-par performance as the MSA-informed model, which suggests that non-MSA features provide excellent distinguishing power for pathogenic and neutral mutations. ALPHAMISSENSE provides a good evaluation for pathogenic mutations but lower specificity for neutral mutations, and achieves the third best MCC. ALPHAMISSENSE may show inflated estimations for metrics based on confusion matrix, since about 8% mutations labeled as 'ambiguous' that are not included. ALPHAMISSENSE has lower AUPRC and AUROC than others, which suggests that it is not robust under other thresholds. EVE exhibits the on-par performance as ALPHAGMUT in recall, specificity, and MCC. However, EVE provides limited predictions, as only 82.8% (1,657/2,000) pathogenic mutations and 55.2% (1,104 by 2,000) neutral mutations can be predicted. Therefore, AUPRC, AUROC, precision, accuracy and F-1 of EVE are biased and not fair to compare. GMVP and POLYPHEN-2 exhibit similar performance that they provide excellent recall, but significantly lower specificity values, which suggests that these two methods may provide more false positives (false pathogenic) in real-world millions of mutations, and increase difficulty to identify true pathogenic mutations.

Besides the performance comparison with these state-of-the-art methods, We also utilize the same sets of feature and dataset to train a random forest model and a multilayer perceptron (MLP) as the baseline models. The random forest has been used in multiple mutation effect predictors and shows good performance [30, 31, 22, 42, 32]. Random forest and MLP intrinsically lack the capability to capture the neighboring impact from spatially relevant residues in the training process. We utilize the same subsets of training and test dataset, and obtain the results in the table 2 Such results prove the effectiveness of the neighboring impact captured by the HOSU and message passing of graph neural network.

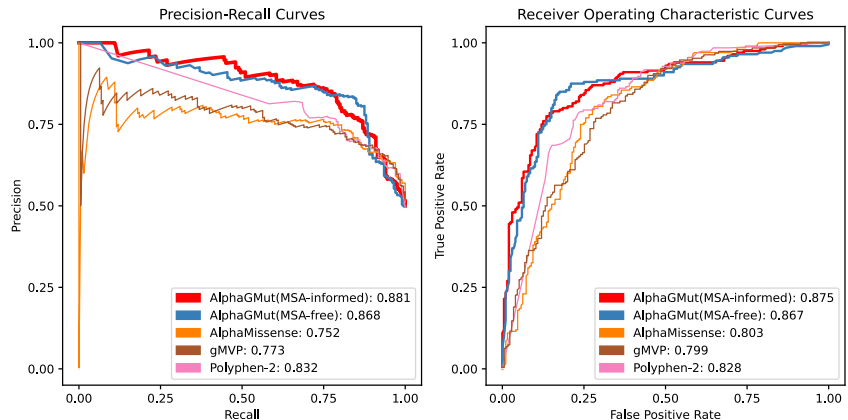

Figure 2: Left) The Precision-Recall curve on the test dataset. Right) The ROC curve on the test dataset.

### 4.2 Better Generalization

Conventional methods usually provide good prediction coverage as they capture features of the mutation site individually. Several recent methods may require the formulation of complex features such as co-evolution signal from deep alignment, which limits their predictability in those domains without sufficient sequence-aware information. We propose that our rationale-driven graph neural network model computes the essential physics-informed features, and intrinsically it does not require deep knowledge of sequence, and should achieve better generalization.

To obtain a fair comparison, we report the ability to predict mutation saturation of 2,267 human proteins with full-length PDB structures in high quality. Our model ALPHAGMUT MSA-free is able to predict mutation effects for 2,266 proteins with all mutation patterns. While EVE exhibits a significantly lower prediction coverage, which only provides predictions on 568 out of 2,267 proteins.

Table 2: Averaged 10-fold Cross-validation Performance Metrics for Alternative Models

|  | ALPHAGMUT MSA-informed | random forest MSA-informed | MLP MSA-informed |
|---|---|---|---|
| AUPRC | **0.878** | 0.742 | 0.856 |
| AUROC | **0.886** | 0.812 | 0.863 |
| Recall | 0.836 | **0.878** | 0.832 |
| Specificity | **0.824** | 0.746 | 0.712 |
| Precision | **0.827** | 0.776 | 0.743 |
| Accuracy | **0.830** | 0.812 | 0.772 |
| F-1 | **0.831** | 0.823 | 0.784 |
| MCC | **0.661** | 0.629 | 0.548 |
| Completeness | 1 | 1 | 1 |

## 5 Conclusion and Future Directions

ALPHAGMUT introduces the novel alpha-shape graph, accurately representing residue connections in 3D structures. It computes rationale-driven features as graph attributes, efficiently guiding the graph neural network to learn the functional impacts of missense mutations and achieve favorable performance. ALPHAGMUT also holds additional advantages: it performs well in MSA-free settings and requires feasible computation hardware in a short training period efficiently. In conclusion, ALPHAGMUT has better performance, interpretability, and generalization.

However, several implementations may further improve the model performance. First, our model formulates the graph with limited edge attributes as it only records the connectivity. We may further consider in-depth edge attributes, such as the bond strength of atomic interaction as edge attribute. Second, we build the message passing according to the HSOU identified by the 4-layer BFS. While residues forming large pockets and voids may not be detected by the 4-layer BFS. Therefore, a better strategy to define the HOSU is required for a more effective message-passing process. Third, our alpha shape graph is built upon an alpha value corresponding to a water molecule at $1.4\,\text{Å}$. We may further assign the alpha values corresponding to binding ligands specific to proteins, which in turn provide protein-ligand specific graphs and may increase the accuracy of graph representation.

## 6 Acknowledgement

We will provide funding information.